\documentstyle[12pt,psfig,epsfig]{article}
\begin{document}

\begin{titlepage}

\title{Obtaining single stimulus evoked potentials with Wavelet Denoising}

\vspace{2cm}
\author{R. Quian Quiroga \\
{\sl John von Neumann Institute for Computing,}\\
{\sl Forschungszentrum J\"ulich,}\\
{\sl D - 52425 J\"ulich, Germany}}
\maketitle
\vspace{3cm}
\hspace{0.3cm}   

\vspace{0.5cm}
\noindent
corresponding author: tel/fax: +49 2461 612302/2430 \\
e-mail: R.QuianQuiroga@fz-juelich.de

\newpage
\begin{abstract}
  We present a method for the analysis of electroencephalograms (EEG).
  In particular, small signals due to stimulation, so called evoked
  potentials, have to be detected in the background EEG. This is
  achieved by using a denoising implementation based on the wavelet
  decomposition.  
  
  One recording of visual evoked potentials, and
  recordings of auditory evoked potentials from 4 subjects
  corresponding to different age groups are analyzed. We find 
  higher variability in older individuals. Moreover, since the evoked
  potentials are identified at the single stimulus level (without need
  of ensemble averaging), this will allow the calculation of better
  resolved averages. 
  Since the method is parameter
  free (i.e. it does not need to be adapted to the particular
  characteristics of each recording), implementations in clinical
  settings are imaginable.
\end{abstract}

\end{titlepage}

\newpage
\section{Introduction}

Evoked potentials (EP) are the alterations of the ongoing
electroencephalogram (EEG) due to stimulation (e.g. tone, light flash,
etc).  They are time locked to the stimulus and they have a
characteristic pattern of response that is more or less reproducible
under similar experimental conditions \cite{basar80,regan}. In order
to study the response of the brain to different tasks, sequences of
stimuli can be arranged according to well defined paradigms.  This
allows the study of 
different sensitive or cognitive functions, states, pathologies, etc,
thus making the EPs an invaluable tool in neurophysiology.

Due to the low amplitudes of evoked potentials in comparison with the ongoing
EEG, they are hardly seen in the original EEG signal, and therefore 
several trials (i.e. data segments including the pre- and
post-stimulus activity) are averaged in order to enhance the evoked
responses.  Since EPs are time locked to the stimulus, their
contribution will add while the ongoing EEG will cancel (see
\cite{locking} for a quantitative study of time locking and amplitude
increases of the EPs in comparison with the spontaneous EEG).  However,
when averaging, information related with variations between the single
trials is lost.  This information could be relevant in order to study
behavioral and functional processes, habituation, refractoriness, and
it could also help to identify pathologies when the information from
the average evoked potential is not clear.  Moreover, in many cases a compromise
must be made when deciding on the number of trials in 
an experiment.  If we take a large number of trials we optimize the
EP/EEG ratio but if the number of trial is too large, then we could
deal with effects such as tiredness, which eventually corrupts the
average results.  This problem can be partially solved by taking
sub-ensemble averages (i.e. consecutive averages of a few single
sweeps). However, in many cases the success of such procedure is
limited, especially when not many trials can be obtained or when 
characteristics of the evoked potentials change from trial to trial.

Several methods have been proposed in order to filter averaged EPs
(see \cite{lopes}).  The success of such methods would imply the need
of less number of trials and would eventually allow the extraction of
{\it single trial evoked potentials} from the background EEG.
Although averaging has been used since the middle 50's, up to now none
of these attempts has been successful in obtaining single trial EPs,
at least in a level that they could be applied to different type of
EPs and that they could be implemented in clinical settings.  Most of
these approaches involves Wiener filtering \cite{walter,doyle} (or a
``minimum mean square error filter" based on auto and cross-correlations
in \cite{mcgillem}) and have the common drawback of considering the
signal as a stationary process.  Since EPs are transient responses
related with specific time and frequency locations, such
time-invariant approaches are not likely to give optimal results. For
this reason, de Weerd and coworkers \cite{deweerd1,deweerd2}
introduced a time-varying generalization for filtering averaged evoked
potentials.  The time-variant Wiener filter they proposed is clearly
more suitable for the analysis of evoked potentials but with the
caveats that such filter bank implementation does not give a perfect
reconstruction, and that it is based on the Fourier Transform
(therefore the signal being decomposed in bases of sines and cosines
with the drawbacks that this imposes, as we will describe later).

These limitations, as well as the ones related with time-invariant
methods can be solved by using the wavelet formalism.  The Wavelet
Transform is a time-frequency representation proposed first in
\cite{grossmann}, that has an optimal resolution both in the time and
frequency domains and has been successfully applied to the study of
EEG-EP signals \cite{wavalpha,schiff,demiralp,bertrand}.  The
objective of the present study is to follow an idea originally
proposed in \cite{bartnik}, and to present a very straightforward
method based on the Wavelet Transform to obtain the evoked responses
at the single trial level.  The key point in the denoising of evoked
potentials is how to select in the wavelet domain the activity
representing the signal (the EPs) and then eliminate the one related
with noise (the background EEG).  In fact, the main difference between
our implementation and previous related approaches is in the way that
the wavelet coefficients are selected. Briefly, such choice should
consider latency variations between the single trial responses and it
should not introduce spurious effects in the time range where the EPs
are expected to occur. In this respect, the denoising implementation
we propose will allow the study of variability between single trials,
an information that could have high physiological relevance.

The paper is organized as follows: In
section \ref{sec-details} the data sets to be analyzed are described.
Section \ref{sec-wav} gives the mathematical background of the Wavelet
Transform and the multiresolution decomposition. The implementation of
the denoising of EPs is presented in section \ref{sec-denoising},
together with a discussion of the advantages of the present
implementation in comparison with previous approaches. The application
of the method to visual and auditory EPs is shown in sections
\ref{sec-vep} and \ref{sec-aep}, respectively. Finally, in section
\ref{sec-conclusion} the conclusions are drawn.

\newpage
\section{Details of the data}
\label{sec-details}

We will study evoked potential recordings obtained with an oddball paradigm upon two
different stimulus modalities: visual evoked potentials (VEP) and
auditory evoked potentials (AEP). The robustness of the method is
stressed by the fact that although an (almost) identical
implementation was used for both data sets, these were taken in
different laboratories and under different recording settings.

\subsection{Visual evoked potentials}

Visual evoked potentials from one normal subject were obtained by
using a checkerboard light pattern.  Two different visual stimuli were
presented in a pseudo-random order (oddball paradigm): $75\%$ of the
stimuli were the so called ``non-target" (a color reversal of the
checks) and the other $25\%$ were the deviant stimuli or ``target"
(also a color reversal but with a half check displacement of the
pattern).
The subject was instructed to ignore the
non-target stimuli and to count the number of appearances of the
target ones (see \cite{wavalpha} for more details on the experimental
setup).
Scalp recordings were obtained from
the left occipital (O1) electrode (near to the location of the visual
primary sensory area) with linked earlobes reference.  
Sampling rate was $250$ Hz and after bandpass filtering
in the range $0.1-70$Hz, $\sim 2sec$ of data (256 data pre- and 256
data post-stimulation) were saved on a hard disk.  Inter-stimulus
intervals varied randomly between $2.5-3.5$ sec. 
The recording session consisted of 200 stimuli presentations. From a total of
50 target stimuli, after rejection of trials with artifacts
(contamination of the recording with spurious activity; e.g.
blinking) 30 were selected for further analysis. 

Two evoked responses can be observed with this paradigm (see average
response on the uppermost left plot in figure \ref{fig:coeff}): 
First, a sensory related positive peak at about 100 ms after stimulation
(P100)  followed by a negative rebound (N200), that appear
both upon non-target and target stimuli. 
Second, a positive peak at about 400-500 ms after stimulation (P300)
appearing only upon target stimuli and related with the cognitive
process of recognizing these stimuli as deviant.

\subsection{Auditory evoked potentials}

Evoked potentials of two young (37 and 45 years) and two old (70 and 74 years) normal subjects
 were obtained in an eyes open condition from a
scalp central electrode (Cz) with linked earlobes reference.  As in the case
of VEP an oddball paradigm was used, the non-target stimuli being
tones of $1000$ Hz and the target ones being tones of $500$ Hz.  The
inter-stimulus interval was $2sec$.  In each session, 100 stimuli
(75 non-target and 25 target) were presented and after artifact
rejection between $40-50$ non-target and between
$11-16$ target stimuli were selected for further analysis.
For each trial, after digital filtering in the range $1-70$ Hz, $1sec$
of data was saved on a hard disc (from $0.2sec$ pre- to $0.8sec$
post-stimulation).  Sampling rate was $204.8$ Hz and the data of
each trial was extended with symmetric border conditions in order to
have a number of 256 data points per trial.

As with VEP, in the case of AEPs two main responses appear: 
the N100-P200 peaks (polarity is reversed in comparison to VEP) and a
P300 response (but appearing earlier than in the VEP case, at about
300ms after stimulation).

\newpage
\section{Wavelet Transform}
\label{sec-wav}

\subsection{From Fourier to Wavelets}

The Fourier Transform of a given signal is defined as its inner
product with complex exponential (sines and cosines) functions of
different frequencies.  It allows a better visualization of the
periodicities of the signal, especially when several frequencies are
superposed.  However, Fourier Transform gives no information about the
time location of these periodicities and it further requires
stationarity of the signal.

By tapering (``windowing") the complex exponential {\it mother
  functions} of the Fourier Transform, the {\it Short Time Fourier
  Transform} (STFT) gives a time evolution of the frequencies that can
be obtained just by sliding the windows throughout the signal.  The
STFT gives an optimal time-frequency representation, but a critical
limitation appears when windowing the data due to the uncertainty
principle (see \cite{chui}).  If the window is too narrow, the
frequency resolution will be poor, and if the window is too wide, the
time localization will be not so precise.  Data involving slow
processes will require wide windows while a narrow window will be more
suitable for data with fast transients (high frequency components).
Then, owing to its fixed window size, the STFT is not suitable for
analyzing signals involving different range of frequencies, as in the
case of EPs.

Grossmann and Morlet \cite{grossmann} introduced the Wavelet Transform in order
to overcome this problem. 
The main advantage of wavelets is that they have a varying window
size, being wide for low frequencies and narrow for the high ones,
thus leading to an optimal time-frequency resolution in all the
frequency ranges \cite{chui}.
Furthermore, owing to the fact that windows are adapted to the
transients of each scale, wavelets do not need stationarity.\\
 
The {\it Continuous Wavelet Transform} (CWT) of a signal $x(t) \in L^2({\cal R})$
is defined as the inner product between the signal and the wavelet
functions $\psi_{a,b}( t )$

\begin{equation}
W_{\psi} x(a,b)~\equiv C_{a,b} ~=~\langle x(t), \psi _{a,b}( t ) \rangle
\label{eq:wt}
\end{equation}

\noindent
where $C_{a,b}$ are the wavelet coefficients and $\psi_{a,b}( t )$ are
dilated (contracted) and shifted versions of a unique {\it wavelet
  function} $\psi(t)$

\begin{equation}
\psi _{a,b}( t )~=~|a|^{-1/2} ~ \psi \left( \frac{t-b}a\right)
\label{eq:mother}
\end{equation}

\noindent
($a,b$ are the scale and translation parameters, respectively). The
CWT gives a decomposition of $x(t)$ in different scales, tending to be
maximum at those scales and time locations where the wavelet best
resembles $x(t)$. Contracted versions of $\psi _{a,b}( t )$ will match
high frequency components of $x(t)$ and on the other hand, dilated
versions will match low frequency oscillations.  

The CWT maps a signal of one independent variable $t$ onto a function
of two independent variables $a,~b$. This procedure is redundant and
not efficient for algorithm implementations. In consequence, it is
more practical to define the Wavelet Transform only at discrete scales
$a$ and discrete times $b$ by choosing the set of parameters $\{ a_j =
2^{j}, b_{j,k} = 2^{j} k \}$, with $j, k \in {\cal Z}$. We obtain then
the discrete wavelet family

\begin{equation}
\psi_{j,k}( t )~=~2^{-j/2}~\psi (~2^{-j}~t~-~k~) 
\quad \quad \quad j,~k~\in~{\cal Z} \ ,
\label{eq:dyadic}
\end{equation}

\noindent
that under appropriate conditions \cite{chui} forms a basis of
$L^2({\cal R})$, each wavelet function having a good localization in
the time and frequency domains. In analogy with eq.(\ref{eq:wt}) we
define the {\it Dyadic Wavelet Transform} as

\begin{equation}
W_{\psi} x(j,k)~\equiv C_{j,k} ~=~\langle x(t), \psi_{j,k}( t )
\rangle  
\label{eq:dwt}
\end{equation}

\noindent
For well defined wavelets it can be inverted, thus giving the
reconstruction of $x(t)$

\begin{equation}
x(t) ~=~ \sum_{j,k} C_{j,k} ~ \hat{\psi}_{j,k} (t)  
\quad \quad \quad j,~k~\in~{\cal Z} 
\label{eq:idwt}
\end{equation}

\noindent
where $\hat{\psi}$ is the dual of ${\psi}$ (in case of orthogonal
wavelets, $\hat{\psi}$ and ${\psi}$ are identical).

\subsection {B-Splines wavelets}
\label{sec-splines}

Another advantage of the Wavelet Transform over Fourier based methods
is that the functions to be matched with the signal are not
necessarily sinusoidal ones (or modulated sinusoidal in the case of
the STFT). 
In fact, there are many different functions suitable as wavelets, each
one having different characteristics that are more or less appropriate
depending on the application.
Irrespective of the mathematical properties of the wavelet to choose, a basic
requirement is that it looks similar to the patterns we want to
localize in the signal. 
This allows a good localization of the structures of interest in the
wavelet domain and moreover, it minimizes spurious
effects in the reconstruction of the signal via the inverse Wavelet
Transform (\ref{eq:idwt}). 

\begin{figure}[c]
\begin{center}
\epsfig{file=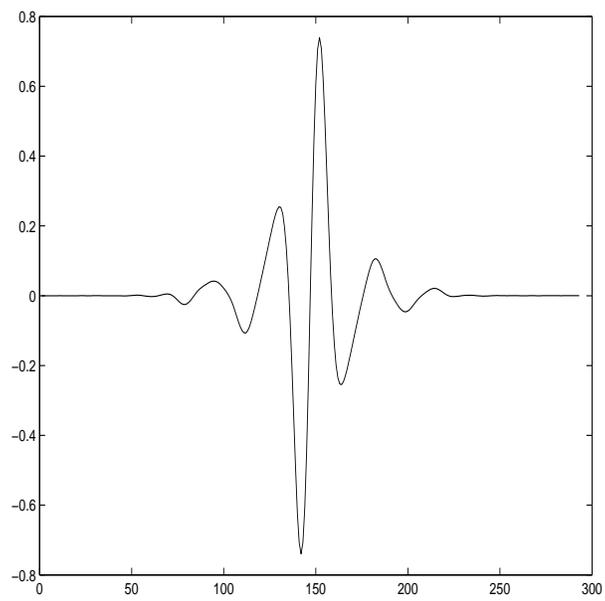,height=8cm,width=8cm,angle=0}
\end{center}
\caption{Quadratic B-Spline wavelet function.}
\label{fig:spline}
\end{figure}

For this study, we choose quadratic biorthogonal B-Splines \cite{cfd}
as mother functions (see Fig.\ref{fig:spline}) due to their similarity
with the evoked responses.  B-Splines are piecewise polynomials that
form a base in $L^2$ \cite{chui,unser}.  We remark the following
properties that make them optimal in signal analysis (see
\cite{unser,chui,cfd} for details): they are (anti-) symmetric, smooth, they
have a nearly optimal time-frequency resolution and they have compact
support.

\subsection{Multiresolution decomposition}
\label{sec-mult}

The information given by the dyadic Wavelet Transform can be organized
according to a hierarchical scheme called multiresolution analysis
\cite{mallat}. If we denote by $W_j$ the subspaces of $L^2$
generated by the wavelets $\psi_{j,k}$ for each level $j$, the space
$L^2$ can be decomposed as a direct sum of the subspaces $W_j$,

\begin{equation}
\label{eq:l2}
L^2 = \sum_{j \in {\cal Z}} W_j 
\end{equation}

\noindent
Let us define the subspaces

\begin{equation}
\label{eq:v}
V_j = W_{j+1} \oplus W_{j+2} \oplus  \ldots   \  \  \  \  j \in {\cal Z}
\end{equation}

\noindent
The subspaces $V_j$ are a {\it multiresolution approximation} of $L^2$
and they are generated by scalings and translations of a single
function $\phi_{j,k} = \phi(~2^{-j}~n~-~k~)$ called the {\it scaling
  function} (see proof in \cite{mallat}).  Then, for the subspaces
$V_j$ we have the complementary subspaces $W_j$, namely:

\begin{equation}
\label{eq:subspaces}
V_{j-1} = V_j  \oplus  W_j   \ \ \ \  j \in {\cal Z} 
\end{equation}

\vspace{1cm} Let us suppose we have a discretely sampled signal $x(n)
\equiv a_0 (n)$ with finite energy.  We can successively decompose it
with the following recursive scheme

\begin{equation}
a_{j-1}(~n~)~= ~a_j (~n~)~+~d_j (~n~)  
\label{eq:bases}
\end{equation}

\noindent
where the terms $a_j (n) = \sum_k~a_{j-1} (k)~\phi_{j,k}(n)~
\in V_j$ give the coarser representation of the signal and $d_j (n) =
\sum_k~a_{j-1}~\psi_{j,k}(n)~\in W_j$ give the details for each
scale $j = 0, 1, \cdots, N$.  For any resolution level $N > 0$, we can
write the decomposition of the signal as

\begin{equation} 
x(n) \equiv a_0 (n) ~=~d_1(n) + d_2(n) + \ldots + d_N(n) + a_N (n)           
\label{eq:multirres}               
\end{equation}

This method gives a decomposition of the signal that can be
implemented with very efficient algorithms\footnote{
The computing time (${\cal O} (M)$,  $M$: number of data
points) is even faster than the one of the fast
Fourier Transform (${\cal O} (M \cdot \log M))$.} due to the recursiveness of
the decomposition (eq.(\ref{eq:bases})). Moreover, Mallat
\cite{mallat} showed that each detail ($d_j$) and approximation signal
($a_j$) can be obtained from the previous approximation $a_{j-1}$ via a
convolution with (FIR or truncated IIR) high-pass and low-pass
filters, respectively. 

De Weerd and coworkers \cite{deweerd1,deweerd2} filtered averaged
evoked potentials by using octave filter banks with the bandwidth of
each filter being proportional to their center frequency.  This
approach is in principle similar to the one obtained with wavelets but
it has some disadvantages. First, it repeatedly makes use of the
Fourier Transform, therefore being forced to decompose the signal in bases of sinus
and cosinus.  
On the contrary, with wavelets we have the advantage of choosing the mother wavelet from a
collection of available functions with properties that will be more or
less suitable depending on the application (see previous section).
Second, as already remarked by Bertrand et al \cite{bertrand} the
filter bank implementation of de Weerd does not allow a perfect
reconstruction of the signal \footnote{ However we should remark that
  the multiresolution decomposition gives a perfect reconstruction
  (disregarding border effects) because the mother functions
  $\psi_{j,k}( t )$ of eq.(\ref{eq:dyadic}) form a basis of $L^2({\cal
    R})$.  This is not to be confounded with the requirement of
  orthogonality as claimed in \cite{bertrand}.}.

The gray curves in Fig.\ref{fig:coeff} show the decomposition of an
averaged (over 30 trials) visual evoked potential.  In this and all
other cases studied we used a $5$ levels decomposition, thus having
$5$ scales of details ($d_1$ to $d_5$) and a final approximation
($a_5$).  On the left side we plot the wavelet coefficients and on the
right side the actual components/decomposition. The sum of all the
reconstructions gives again the original signal (gray curve of the
uppermost right plot).  The lower levels give the details
corresponding to the high frequency components and the higher levels
the ones correspond to the low frequencies.

\newpage
\section{Denoising of evoked potentials}
\label{sec-denoising}

\subsection{Implementation of the method}

As proposed by Donoho \cite{donoho}, the conventional definition of
denoising implies a thresholding criterion in the wavelet domain.
The signal is recovered from noisy data just by setting to zero
those wavelet coefficients below a certain threshold ({\it hard
  denoising}) or with the use of a smoother transformation ({\it soft
  denoising}).  However, this procedure is not optimal for recovering
the evoked potentials because these ones are of the order or even smaller
than the background EEG. Therefore, instead of using a thresholding
criterion, we implemented a denoising based on the specific time and
frequency localizations of the evoked responses.

\begin{figure}
\begin{center}
\epsfig{file=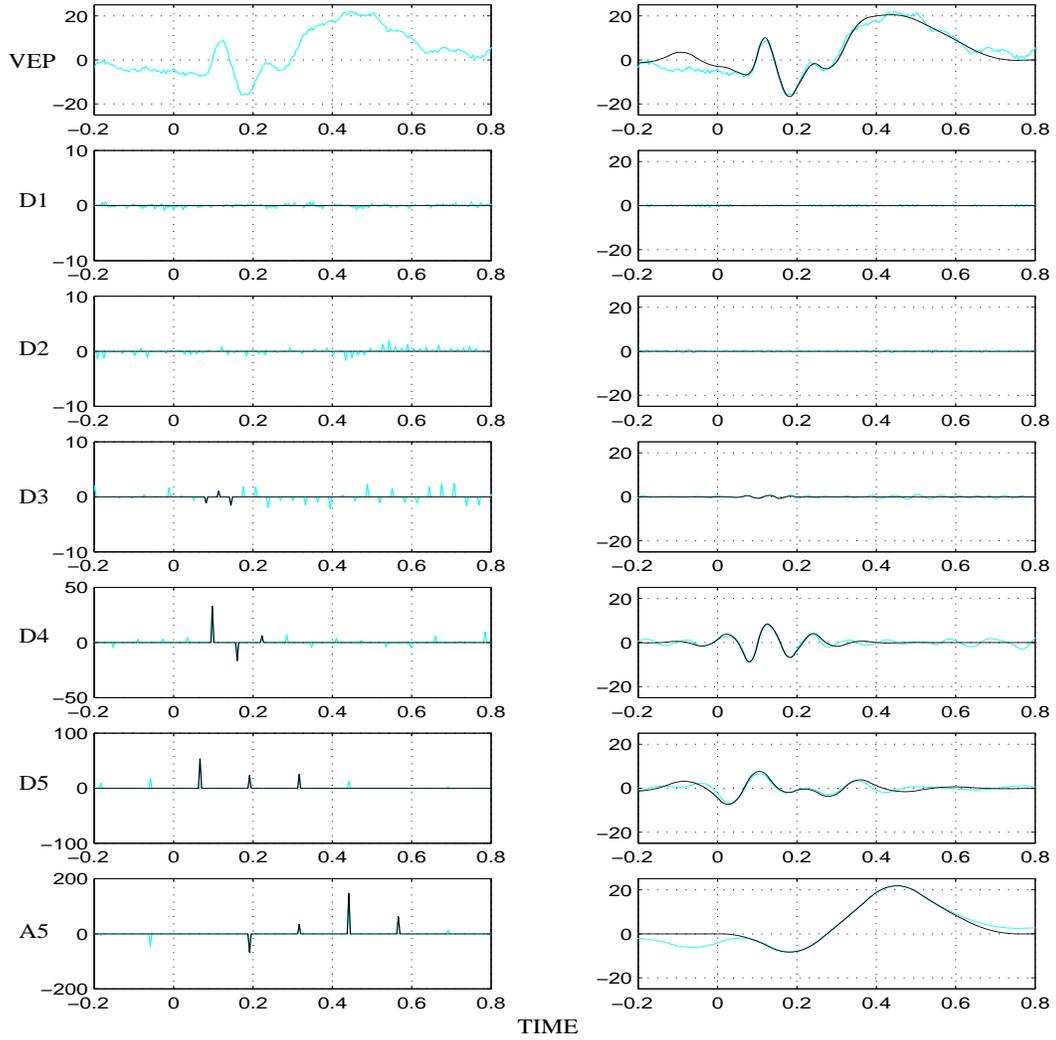,height=14cm,width=14cm,angle=-90}
\end{center}
\caption{Multiresolution decomposition and reconstruction of an
  averaged visual evoked potential (target stimuli). Gray curves: original decomposition
  and reconstruction; black curves: denoised decomposition and
  reconstruction. $d_1-d_5$ are the details at different scales and
  $a_5$ is the last approximation. Note how the denoised
  reconstruction filters the noisy activity contaminating the original
  average.}
\label{fig:coeff}
\end{figure}

Note in Fig.\ref{fig:coeff} that the P100-N200 response is correlated
mostly with the first post-stimulus coefficients in the details $d_4 -
d_5$ and the P300 is mainly correlated with the coefficients at about
$400-500$ms in $a_5$.  In consequence, a
straightforward way to avoid the fluctuations related with the ongoing
EEG and get only the peaks of interest, is just by equaling to zero
those coefficients not correlated with the EPs.

The black traces in the left side of fig.\ref{fig:coeff} show the
coefficients kept for the reconstruction of the P100-N200 and P300
responses and the black curves on the right side show the
contributions of each level obtained by eliminating all the other
coefficients.  Note that in the final reconstruction of the averaged
response (black curve in the uppermost right plot) background EEG
oscillations are cancelled.  We should remark that this is usually
difficult to achieve with a Fourier filtering approach (especially in
averages of less number of trials) due to the different time and
frequency localizations of the P100 and P300 responses, and
overlapping frequency components of these peaks and the ongoing EEG.
In this context, the main advantage of Wavelet denoising over
conventional filtering is that we can select different ``time windows"
for the different scales. Advantages of wavelets over Fourier for
simultaneously filtering early and middle latency (up to 50 ms after
stimulation) averaged auditory EPs were described by Bertrand et al.
\cite{bertrand}.
Moreover, these authors proposed a ``wavelet based"\footnote{ 
  Instead of the usual Wavelet Transform, 
  Bertrand and coworkers used an analogous transform (the wavelet functions
   being not anymore dilated and translated versions of a unique wavelet)
  in order to avoid border problems in the reconstruction. However,
  such correction seems to be unnecessary in general because 
  border problems can be easily avoided just by having enough pre- and
  post-stimulus data as is usually the case in EP recordings.} 
post-averaging Wiener
filtering for cleaning averaged evoked potentials.

Since as we mentioned in the introduction, evoked potentials are 
 activity time locked to the stimulation, we could use this feature
to reconstruct the contribution of the single trials to the averaged
EPs.  This will allow the visualization of the EPs at the single trial
level as we will show in the next section.

Let us remark a critical point when implementing the denoising
of the EPs. This is the choice of which coefficients to
keep and which to eliminate. 
On one hand, choosing a wide range of scales (``frequency window") allows a
better reconstruction of the morphology of the EPs (again we remark
that the selection of an appropriate wavelet function plays an
important role in this respect). Also, choosing a wide ``time window" of
coefficients makes the method more sensitive to latency differences
(jitters) between trials (in the extreme case of keeping one single
coefficient, the denoised signal will be just the wavelet function
with its amplitude proportional to the coefficient).
On the other hand, if we choose a wide (``conservative") range of coefficients 
we would not enough eliminate the background EEG activity in order to 
recognize the EPs (in the extreme case of keeping all coefficients we
will just reconstruct the original signal).
In this respect we propose to use test signals, such as an spontaneous
EEG (or a pre-stimulus EEG segment as we will show in
fig.\ref{fig:cont-eeg}) in order to check for eventual
spurious interpretations due to an unfortunate selection of the coefficients.

We heuristically found the selection of those coefficients remarked in black
traces in Fig.\ref{fig:coeff} an optimal compromise between EP
resolution and sensitivity to variations between trials. 
As we will show in the next sections, they allow the visualization of
the single trial EPs and also they cover a reasonable time and scale
ranges where the EPs are physiologically expected to occur.

We should also mention that the wavelet coefficients to be
kept could be smoothed by using, e.g. {\it soft thresholding}
\cite{donoho} in order to decrease border effects. Although the {\it
  hard thresholding} we used can introduce spurious border
fluctuations (see for example in Fig.\ref{fig:coeff} the positive
deflection in the denoised signal between $-0.2$ and $0$sec), these are
outside the time range of physiological interest of the evoked
responses.

In summary the method consist in the following steps:
\begin{enumerate}
\item The averaged EP is decomposed by using the wavelet
  multiresolution decomposition.
\item The wavelet coefficients not correlated with the average evoked
  potential (but also considering a time range in which single trial
  EPs are expected to occur) are identified and set to zero.
\item The inverse transform is applied, thus obtaining a denoised
  signal.
\item The denoising transform defined by the previous steps is applied
  to the single trials.
\item Finally, validity of the results can be checked by applying the
  method to EEG test signals.
\end{enumerate}

We remark that for all the calculations done in this study we
keep the same set of coefficients (with the slight exception that for
auditory evoked potentials we do not keep the last coefficient of level $A_5$ because
auditory P300 responses appear earlier than in the visual case). 
In other words, once the coefficients are chosen (steps 1-3), the method is
parameter free and does not need to be adjusted for different EEG/EP
ratios, latency variability, number of trials or other differences
between subjects, electrodes, etc.

\subsection{Comparison with previous works}

The criteria for choosing which wavelet coefficients are correlated
with the signal and which ones with noise is the key feature of the
different denoising implementations.  This is in fact our main
difference with previous approaches. As already mentioned, Donoho \cite{donoho} proposed a
thresholding criteria, something not suitable for separating evoked
potentials from the background EEG activity. The aim of those methods using time-variant \cite{walter,doyle}, time-invariant
\cite{deweerd1,deweerd2} and wavelet based \cite{bertrand} Wiener
filtering was to clean averaged EPs rather than to obtain single trial EPs\footnote{
However, since average filtering implies the need of less number of
trials, the idea of single trial reconstruction is implicitly
involved.}. 
Therefore, they did not consider latency variations between trials,
one of the most important features obtained from single trial analysis in
comparison with ensemble averaging.  A similar remark is applicable to
the work of Demiralp et al \cite{demiralp}. These authors correlated
one single wavelet coefficient with the P300 response, then using its sign
for discriminating between trials with and without P300. By selecting
trials in this way, they succeeded in obtaining better averages of the
P300.

Mc Gillen and Aunon \cite{mcgillem} and Bartnik et al \cite{bartnik}, 
instead, aimed at obtaining single trial evoked potentials.  
The first authors proposed a filter based in auto- and
cross-correlations that however, due to its time-invariance, is not
optimal for the analysis of EPs. This is the main explanation for the
spurious differences between the original and the filtered signals
observed in their examples (figs. 3 and 4 in
ref.\cite{mcgillem}).  Bartnik and coworkers introduced a
denoising implementation similar to the one presented by us.  However,
based on correlation and discriminant analysis (using as input the
wavelet coefficients from the average and the single trials) they
proposed an automatic criterion for finding which wavelet coefficients
best distinguishes the EP from the EEG.  In comparison with our
approach, this criterion has the the following drawbacks: First, it is
not appropriate if more than one peak is present, as e.g. in the case
of the P100-N200 and P300 responses to be shown in the next section.
Second, it is in principle not sensitive to variations between trials
(i.e. variations in latency will be cancelled in the average and
consequently, they will be disregarded by the correlation analysis).
Third, a denoising based in such correlation criterion is likely to
alter the morphology of the EPs. In particular, in the example shown
by Bartnik et al the denoised signals look much smoother than the
original ones, in many cases worsening the recognition and time
localization of the EPs (see fig.7 in ref.\cite{bartnik}).

\newpage
\section{Application to Visual Evoked Potentials}
\label{sec-vep}

\begin{figure}
\begin{center}
\epsfig{file=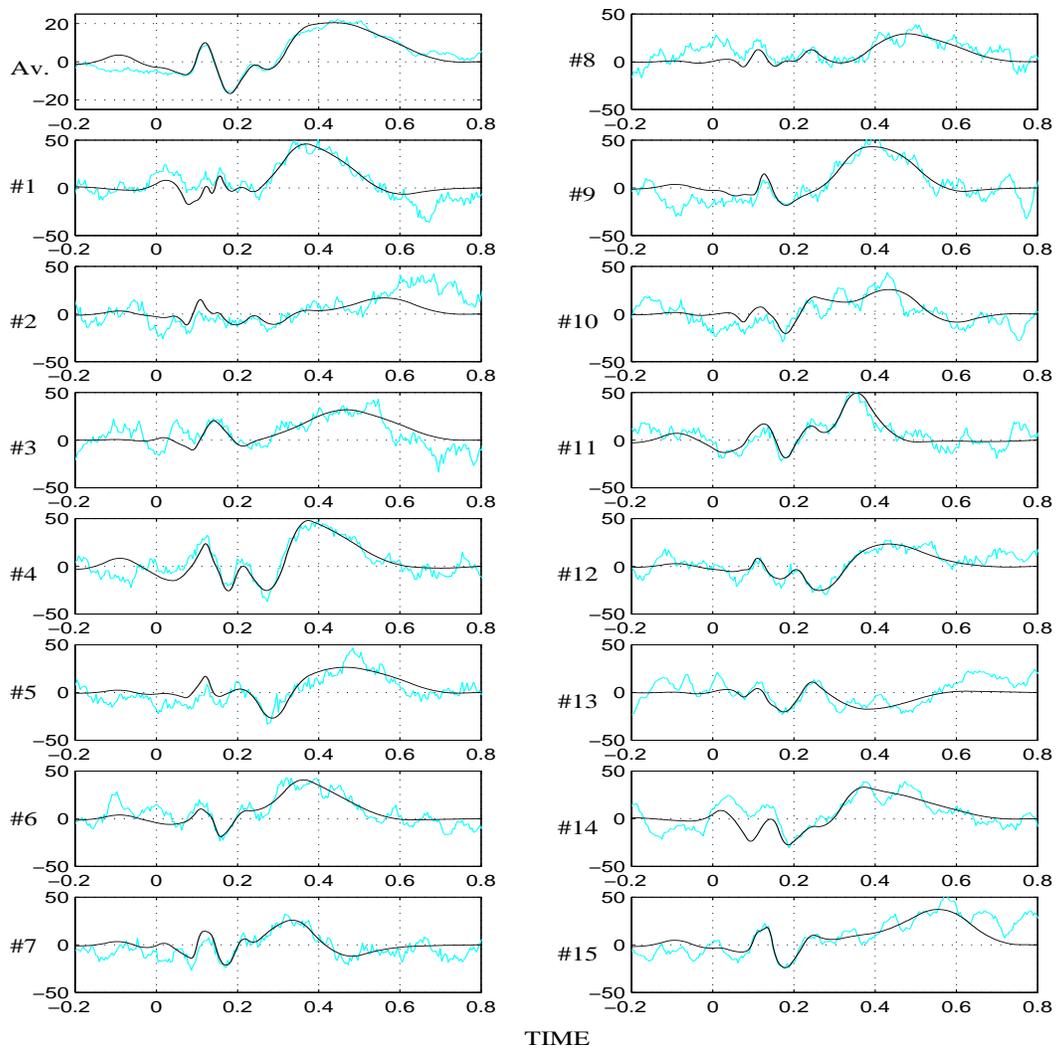,height=14cm,width=14cm,angle=-90}
\end{center}
\caption{Single trials corresponding to the average of the previous
  figure. Grey lines: original data; black lines: denoised data. Note
  how after denoising the evoked potentials (as seen in the average) are
  recognizable in most of the single trials from the background EEG.}
\label{fig:sweeps-ja}
\end{figure}

Figure \ref{fig:sweeps-ja} shows the first $15$ single trials
corresponding to the average of Fig.\ref{fig:coeff}. Note that with
denoising (black curves) we can distinguish the evoked responses
P100-N200 and P300 in most of the trials. Note also that these
responses are not easily identified in the original signal due to
their similarity with the ongoing EEG. We already can observe some
variability between the trials, e.g. in the trials $\#2$,$\#13$ the
EPs are practically not present.
For easier visualization, in Fig.\ref{fig:cont-ja}
we plot the single trial evoked potentials with and without denoising
(left and right side plot, respectively) by using contour plots. In
the denoised plot we observe between $100-200$ms a yellow/red pattern
followed by a blue one corresponding to the P100-N200 peaks. These
responses remain more or less stable during the whole recording
session.  The more unstable and wider yellow/red pattern at about
$400-600$ms correspond to the cognitive related P300 responses.
Noteworthy, all these patterns are more difficult to recognize in the
original signal (left plot).

In order to check that the better visualization obtained after
denoising is not just an artifact of the method, in
Fig.\ref{fig:cont-eeg} we show the results of the same analysis
applied to an spontaneous EEG signal (without EPs). As the EEG signal,
we took for each trial of the VEP recording set the data segments
corresponding to the second previous to stimulation (we assume we can
disregard pre-stimulus effects, such as expectation). In contrast to
the previous figure we only observe some randomly appearing spots that
do not form a coherent pattern as when the EPs are present.

\section{Application to Auditory Evoked Potentials}
\label{sec-aep}

Auditory target and non-target evoked potentials of a typical
subject (427) are shown in Figs.\ref{fig:cont-427t}  and \ref{fig:cont-427}, respectively.
In both figures we observe a blue pattern at about 100ms after
stimulation corresponding to the N100, followed by a yellow/red
pattern corresponding to the P200. Upon target stimuli we can also
identify a yellow pattern at about 300ms corresponding to the P300
responses. The P300 is better defined in the first 5 trials, later
it appears more spread and unstable. 

We further study the variability of the N100 peaks upon non-target
stimuli (variability of P300 upon target stimuli can be studied in a
similar way).  We first obtained from the denoised responses the
amplitude and latency of the N100 peaks , which where localized as the
maximum in a time window between $50-180$ms after stimulation.  We
remark that this is hard to implement in the original signal due to
the presence of the ongoing EEG.  We then calculated the mean and
standard deviation of the peak amplitude and latency.  The first three
columns of Table \ref{tab:averages} show the results for the four
subjects studied and the control EEG signal (without EPs) already
described in Fig.\ref{fig:cont-eeg}.  The two different groups of
subjects (young and old) can not be distinguished from the mean
amplitudes or latencies (with exception of the fourth subject, which
will be discussed later). However, we observe a higher variability both
in amplitude and latency for the older subjects. These variations
are even larger for the control EEG signal.  We remark that our
present goal is to illustrate how we can obtain information not seen
in the EP averages rather than making any physiological conclusion. In
this context, a detailed quantitative analysis of the variability
between single trials will be reported elsewhere with a larger
database of subjects and statistical validation.

We also calculated the cross-correlations between the single trials and
the average for each subject, thus having a measure of ``how good" the single trials
resemble the average EP.  Then we computed {\it selective averages}
using only those trials whose cross-correlation with the average was larger than
$0.4$ ~\footnote{Similarly, a cross-correlation criteria can be also implemented
  for separating different type of evoked responses as shown in
  \cite{woody}.}. 
Although we will show selective averages for the original and
denoised signals, cross-correlations were only computed with the denoised
ones in order to avoid effects of random correlations of the
background EEG. 
As implemented in \cite{woody} for pre-filtered trials, for the
selected trials we calculated {\it lag corrected averages}, i.e. by
shifting each trial $\Delta_i = t_{av} - t_i$ data points, where
$t_{av}$ is the latency of the average and $t_i$ the one of each
single sweep.  The last two columns of table \ref{tab:averages}
reports the mean cross-correlation values and the number of trials
selected for each subject. As expected, the EEG signal has a lower mean
cross-correlation than the EPs.

In Fig.\ref{fig:averages} we show the averages, the selective averages
and the shift corrected ones for the 4 subjects studied and the
control EEG. On the left plots the averages with the original
signals are shown and on the right plots the ones with the
denoised ones.  We first observe that in the denoised signals the
spontaneous EEG oscillations are cancelled, thus having a clearer
visualization of the EPs.  
For the first three subjects, in both original and denoised signals 
the selective averages show a better definition of the EPs. The same
holds for the shifted averages, especially in the third case due to
its larger latency variation (see Table \ref{tab:averages}). 
In the forth subject as in the case of the EEG, neither the selective
and the shifted averages show a better identification of the EPs in
comparison with the background activity.
Note also that this is expected from their low cross-correlation
values. Again, a more reliable distinction between subjects with a very large
EP variability and control EEGs would require the statistical analysis of a
larger sample, something that it is beyond the scope of the present study.

\newpage
\section{Conclusions}
\label{sec-conclusion}

We presented a method for extracting evoked potentials from the background EEG. It
is not limited to the study of EPs/EEGs and similar implementations
can be used for recognizing transients even in signals with low signal
to noise ratio.  The denoising of evoked potentials allowed the study
of the variability between the single responses, information that could
have a high physiological relevance in the study of different brain
functions, states or pathologies. It could be also used to eliminate
artifacts that do not appear in the same time and frequency ranges of
the relevant evoked responses.  In passing, we showed that the method
gives better averaged evoked potentials due to the high time-frequency resolution of
the Wavelet Transform, this being hard to achieve with conventional
Fourier filters.  Moreover, since trials with ``good" evoked responses
can be easily identified, it was possible to do selective averages or
even jitter corrected ones, with a resulting better definition of the
averaged EPs.  These advantages could significantly reduce the minimum
number of trials necessary in a recording session, something of high
importance for avoiding behavioral changes during the recording (e.g.
effects of tiredness) or, even more interesting, for obtaining EPs
under strongly varying conditions, as with children, or patients
with attentional problems.

The idea of denoising for obtaining single trial evoked potentials
together with a
single example were already presented in \cite{bartnik}. The main
difference with these authors is in
the criteria for selecting the wavelet coefficients. This is crucial
for obtaining an optimal implementation that is physiologically
plausible (e.g. allowing variations between trials) and that
minimizes the presence of spurious effects in the time range of the
evoked responses.

Finally, We should remark that once the wavelet coefficients to be used for 
denoising were selected (according to the evoked responses to focus on),
the procedure for obtaining single trial EPs and for improving the
averaged responses (due to denoising of the averages, to selective
averaging or to jitter corrections) was done fully automatically for
all trials and patients.  Therefore, implementations of the method as
a complementary information to routine EEG/EP analysis are imaginable.

\section*{Acknowledgments}

We are very thankful to Dr. H. Garc\'{\i}a for the AEP data and to
Prof. E. Ba\c{s}ar and Dr. M. Sch\"urmann for the VEP data. We are
also very thankful to Prof. P. Grassberger for stimulating discussions and a
careful reading of this manuscript.

\bibliography{synchro}

\newpage

\begin{table}[h]
\begin{center}
\begin{tabular}[c]{c c c c c}
{\bf Subject} & {\bf Ampl. N100}  & {\bf latency N100} & {\bf
  cross-corr.} & {\bf selected trials} \\ 
{\bf (age)} & {\bf (S.D.)}  & {\bf (S.D.)} &  & \\ 
\hline
427  (45) & 3.80 (4.24) & 127 (2.69) &  0.43 & $28/50$ (46\%)\\ 
594  (37) & 5.09 (5.46) & 117 (2.42) &  0.40 & $25/40$ (52\%)\\
492  (70) & 3.68 (9.52) & 132 (3.61) &  0.40 & $16/50$ (32\%)\\ 
1558 (74) & 1.87 (8.95) & 107 (3.13) &  0.30 & $16/40$ (35\%)\\ 
EEG       & 1.95 (12.00)&  93 (5.93) &  0.14 & $6/30$  (13\%)\\ 
\end{tabular}
\caption{Results of the quantitative analysis of auditory evoked potentials (N100). Note the
  higher variability in the amplitude and latency of the N100 in the
  older subjects. In the last row we include the values for a control
  EEG signal (without evoked potentials).}
\label{tab:averages}
\end{center}
\end{table}

\newpage

\begin{figure}
\begin{center}
\end{center}
\caption{[COLOR] Contour plot of the single trials shown in the
  previous figure. After denoising (rigth plot) it is possible to see the evolution
   of the evoked responses with the trial number. This is hardly
  recognizable from the original signal (left plot).}
\label{fig:cont-ja}
\end{figure}

\newpage

\begin{figure}
\begin{center}
\end{center}
\caption{[COLOR] Contour plot of an ongoing electroencephalogram (without evoked
  potentials) and its denoising. Note
  that in this case no coherent structure is visible in the denoised signal.}
\label{fig:cont-eeg}
\end{figure}


\newpage

\begin{figure}
\begin{center}
\end{center}
\caption{[COLOR] auditory evoked potential ``target trials" of a typical subject (subject:427, age: 45) and the denoised responses. Note in particular the decrease
  of the P300 response after the fifth trial. Again, these patterns
  are hardly recognizable from the original signal.}
\label{fig:cont-427t}
\end{figure}

\begin{figure}
\begin{center}
\end{center}
\caption{[COLOR] auditory evoked potentials ``non-target trials" of the  
   subject of the previous
  figure. As expected, no pattern related with a P300 is seen.}
\label{fig:cont-427}
\end{figure}

\begin{figure}
\begin{center}
\end{center}
\caption{Averaged auditory evoked potentials, selective averages and jitter corrected
 averages for the 4 subjects studied and the spontaneous EEG
 signal. Note how selective and jitter corrected averages increase the
 definition of the evoked potentials (except for subject 1558 where
 changes are comparable to the ones obtained in the control EEG signal. }
\label{fig:averages}
\end{figure}

\end{document}